\documentclass[twocolumn,prl,superscriptaddress,aps]{revtex4-1}

\usepackage{graphicx}
\usepackage{amsmath}
\usepackage{amssymb}
\usepackage{epstopdf}

\begin{document}

\title{Evolution of Magnetic Excitations Across the Metal-Insulator Transition in a Pyrochlore Iridate Eu$_{2}$Ir$_{2}$O$_{7}$}
	

\author{Sae~Hwan Chun,$^{1}$ Bo Yuan,$^{1}$ Diego Casa,$^{2}$ Jungho Kim,$^{2}$ Chang-Yong Kim,$^{3}$ Zhaoming Tian,$^{4}$ Yang Qiu,$^{4}$  Satoru Nakatsuji,$^{4}$ and Young-June Kim$^{1}$}
\affiliation{
	$^{1}$Department of Physics, University of Toronto, 60 St. George Street, Toronto, ON M5S 1A7, Canada\\
    $^{2}$Advanced Photon Source, Argonne National Laboratory, Argonne, IL 60439, U.S.A.\\
	$^{3}$Canadian Light Source, 44 Innovation Blvd., Saskatoon, SK S7N 2V3, Canada\\
	$^{4}$Institute for Solid State Physics, University of Tokyo, Kashiwa 277-8581, Japan		
	}




\begin{abstract}
We report Resonant Inelastic X-ray Scattering (RIXS) study of the magnetic excitation spectrum in a highly insulating Eu$_{2}$Ir$_{2}$O$_{7}$ single crystal that exhibits a metal-insulator transition at $T_{MI}$ = 111(7) K. A propagating magnon mode with 20 meV bandwidth and 28 meV magnon gap is found in the excitation spectrum at 7 K, which is expected in the all-in-all-out (AIAO) magnetically ordered state. This magnetic excitation exhibits substantial softening as temperature is raised towards $T_{MI}$, and turns into highly damped excitation in the paramagnetic phase. Remarkably, the softening occurs throughout the whole Brillouin zone including the zone boundary. This observation is inconsistent with magnon renormalization expected in a local moment system, and indicates that the strength of electron correlation in Eu$_{2}$Ir$_{2}$O$_{7}$ is only moderate, so that electron itinerancy should be taken into account in describing its magnetism.
\end{abstract}


\maketitle

Iridates exhibit a plethora of diverse and interesting electronic phases \cite{BJK, Machida, Matsuhira, YKK, Ueda, Tian, Chu, Liang}; some of which are predicted to be topologically non-trivial due to the strong spin-orbit coupling in this material \cite{Pesin}. One of such topological phases is Weyl semimetal, which was first predicted in pyrochlore iridates \cite{Wan, WK}, in which Ir$^{4+}$ ions are found at the vertices of tetrahedra forming a corner-sharing network. This topological phase can be present in the all-in-all-out (AIAO) magnetic state where all the Ir$^{4+}$ magnetic moments either point in or out of the center of the tetrahedron. In solid state, a Weyl semimetal results from the band touching at special points in momentum space called Weyl nodes, of which projection on the surface Brillouin zone is connected by a Fermi-arc. An important requirement for this type of band structure is the removal of band degeneracy near the Fermi level via breaking either time-reversal or space-inversion symmetry. The latter case is realized in the case of TaAs and related materials \cite{Weng}, in which a Fermi-arc surface state, a hallmark of Weyl semimetal band structure, was observed recently \cite{Xu,Lv}. The situation in pyrochlore iridates in which time-reversal symmetry is broken, is still unsettled.

One of the difficulties is the fact that electron correlation effects are significant in iridates. In fact, Sr$_{2}$IrO$_{4}$ is considered as a Mott insulator, which shares many similarities with parent cuprates \cite{BJK, YKK, JHK} rather than other topological insulators or TaAs, whose band structure is captured by non-interacting calculations \cite{Weng}. Pyrochlore iridates, on the other hand, are expected to exhibit much more itinerant electron character as indicated by its proximity to a metal-insulator transition (MIT) as a function of temperature \cite{Matsuhira, Ueda, Ishikawa} or pressure \cite{Tafti}. Since strong correlation tends to drive the system to Mott insulating regime, weaker intermediate correlation is a requisite for realizing topological semimetal phases in experimental systems \cite{Wan, WK}. Classification of electron correlation in the pyrochlore iridates, therefore, is a crucial first step for understanding topological phase in these materials. However, to date, there has been no consensus regarding the strength of electron correlation in the pyrochlore iridates \cite{Wan, WK, Lee, Sushkov, Nakayama, Ueda02}.

In this letter, we report temperature dependent evolution of magnetic excitation spectrum in Eu$_{2}$Ir$_{2}$O$_{7}$ in order to examine electron correlation. The magnetic excitation spectrum obtained via Resonant Inelastic X-ray Scattering (RIXS) exhibits a dramatic contrast between local and itinerant magnetism. At low temperatures, the excitation spectrum is well described by a propagating magnon mode with 20 meV bandwidth and a 28 meV gap. We find that the magnon gap scales with the magnetic order parameter and disappears at the MIT temperature. Surprisingly, we also find strong suppression of the zone boundary excitation energy with increasing temperature. The rigid shift of magnetic excitation band cannot be explained by thermal magnon renormalization as expected in a local moment system, and points to a substantial change in the electronic structure that accompanies the magnetic transition. This result strongly advocates intermediate electron correlation in Eu$_{2}$Ir$_{2}$O$_{7}$, implying that this pyrochlore iridate is compatible with Weyl semimetal physics.

The RIXS experiment was performed at the 27 ID-B beamline of the Advanced Photon Source. The instrumental setup with overall energy resolution of 28 meV (FWHM) at the incident energy 11.215 keV (near the Ir $L_{3}$ edge) was used to measure magnetic excitations. In order to minimize elastic background intensity, horizontal scattering geometry was used and excitations in the Brillouin zone around \textbf{\textit{G}} = (7, 7, 7) (in the cubic notation) whose scattering angle 2$\theta$ is close to 90 degrees was studied. We also carried out complementary Resonant Magnetic X-ray Scattering (RMXS) experiment at the HXMA beamline of the Canadian Light Source. Single crystals of Eu$_{2}$Ir$_{2}$O$_{7}$ were grown by flux method using the polycrystalline powder and KF flux as described in Ref. \cite{Ishikawa}. Samples from the same source was used in previous RMXS investigation \cite{Sagayama}. These samples are highly insulating unlike the samples used in earlier RIXS investigation \cite{Clancy}.

\begin{figure} 
	\centering
	\includegraphics[width=3.1in]{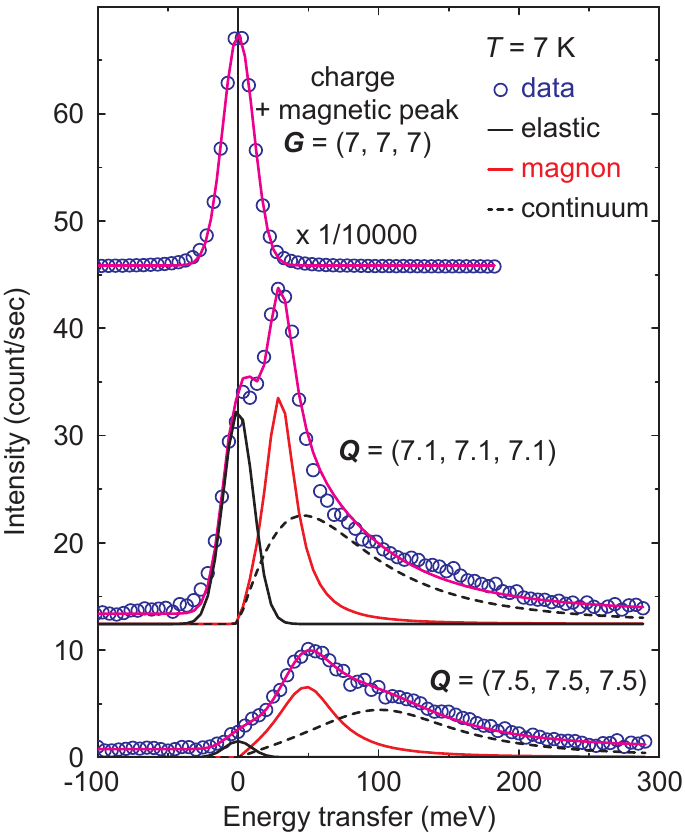}
	\caption{
		The RIXS spectra at \textit{T} = 7 K. The spectrum at (7, 7, 7), the center of the structural and magnetic Brillouin zone, displays the elastic line profile fitted by Gaussian line profile with FWHM = 28 meV (Top). The intensity is reduced by a factor of $10^{4}$ for comparison with the other spectra. The fitting curves for the spectra at \textit{\textbf{q}} = (0.1, 0.1, 0.1) and (0.5, 0.5, 0.5) ($\textit{\textbf{q}} \equiv \textit{\textbf{Q}} - \textit{\textbf{G}}$) comprise the elastic background (solid black), inelastic peak from the magnetic excitation (red), and inelastic background continuum (dashed black) (Middle and Bottom). The spectra are shifted vertically for clarity.
	}
	\label{fig:example}
\end{figure}

Figure 1 shows the RIXS spectra in the energy transfer range below 300 meV obtained at \textit{T} = 7 K. Representative spectra near the zone center \textit{\textbf{q}} = (0.1, 0.1, 0.1) and at the zone boundary (0.5, 0.5, 0.5) show pronounced inelastic peaks at \textit{E} $\approx$ 28 meV and 48 meV, respectively, in addition to the background consisting of sharp elastic peak and a broad feature at the high energy tail of the inelastic peak. We find that the inelastic peak has strong temperature dependence as shown below, indicating its magnetic origin. On the other hand, the origin of the inelastic background extending to 300 meV is not clear, but we speculate that it is due to particle-hole continuum or incoherent multi-magnetic excitation based on the broad linewidth and the absence of noticeable temperature dependence.

To elucidate the dispersion relation of the magnetic excitation, we obtained the RIXS spectra along high symmetry directions in the Brillouin zone as shown in Figs. 2(a-c). In all three directions, the elastic background is greatly reduced as we move away from the $\Gamma$ point, \textbf{\textit{G}} = (7, 7, 7), which is also magnetic Bragg peak position. One can also observe that the inelastic peak intensity decreases and its peak position moves to higher energy. In Fig. 2(d), the intensity is plotted as a function of energy transfer $\omega$ and momentum transfer \textit{\textbf{q}} in pseudo-color scale.

To obtain quantitative momentum dependence, we fit the magnetic inelastic peak at each \textit{\textbf{q}} with a damped harmonic-oscillator model:
\begin{equation}
	\resizebox{.65\hsize}{!}{$S(\textit{\textbf{q}},\omega)=\frac{A(\textit{\textbf{q}})}{1-\exp(- \omega/T)}[\frac{\gamma_{\textit{\textbf{q}}}}{(\omega-\omega_{\textit{\textbf{q}}})^{2}+\gamma_{\textit{\textbf{q}}}^{2}}-\frac{\gamma_{\textit{\textbf{q}}}}{(\omega+\omega_{\textit{\textbf{q}}})^{2}+\gamma_{\textit{\textbf{q}}}^{2}}$]},
\end{equation}
where $\omega_{\textit{\textbf{q}}}$ is the peak position, $\gamma_{\textit{\textbf{q}}}$ is the peak width, and \textit{A}(\textit{\textbf{q}}) is the overall amplitude. We set $\hbar = k_B = 1$ throughout this paper. In addition, the same functional form was used for the broad background, and the elastic background was fitted with the resolution function determined from the spectrum at \textbf{\textit{G}} = (7, 7, 7). Examples of the fitting are shown in Fig. 1. The peak positions ($\omega_{\textit{\textbf{q}}}$) extracted from the fitting are overlaid on top of the intensity map shown in Fig. 2(d). The extracted peak widths ($\gamma_{\textit{\textbf{q}}}$) on the other hand remains almost resolution limited throughout the Brillouin zone, indicating that the magnetic excitation is described as a propagating, long-lived, magnon mode.

\begin{figure} 
	\centering
	\includegraphics[width=3.1in]{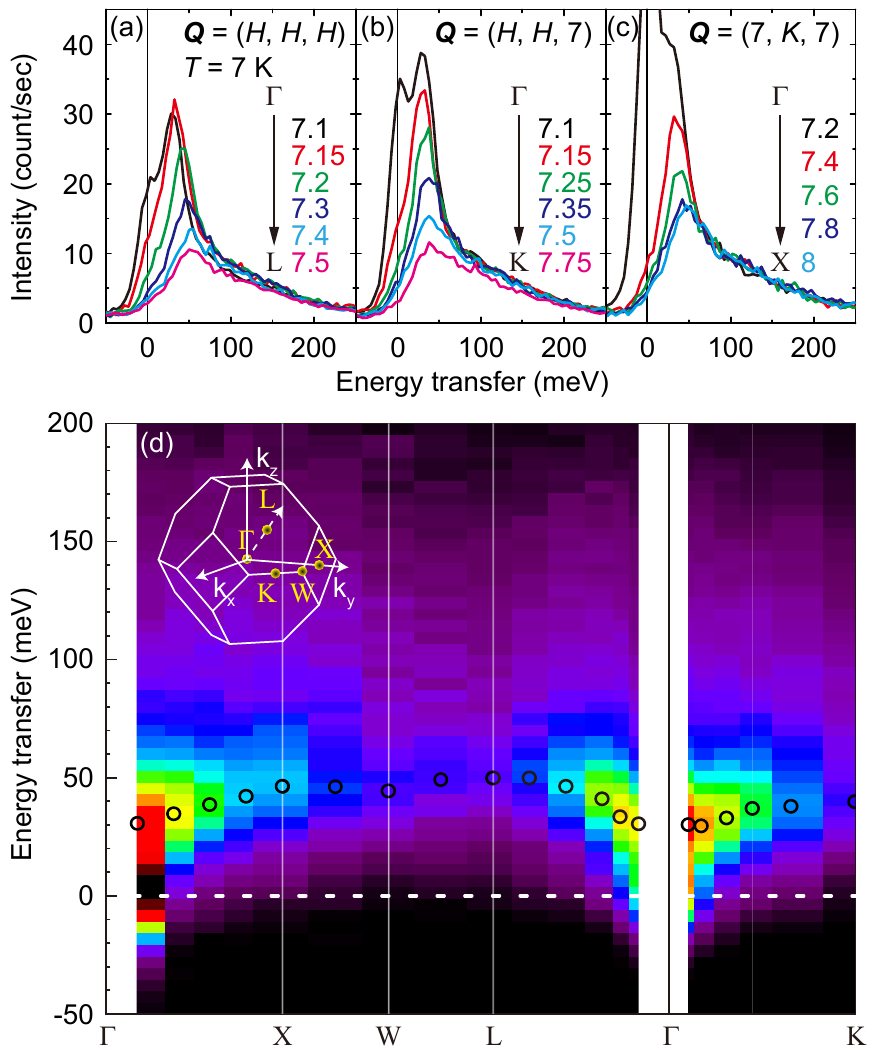}
	\caption{
		The RIXS spectra at \textit{T} = 7 K along the high symmetric directions: the $\Gamma \rightarrow$ L (a), $\Gamma \rightarrow$ K (b), and $\Gamma \rightarrow$ X (c) lines in the Brillouin zone centered at \textit{\textbf{G}} = (7, 7, 7). (d) Intensity map of the RIXS spectra. The open circles denote the $\omega_\textbf{\textit{q}}$ positions of the magnetic excitations. The Inset shows the Brillouin zone and notations for high symmetry positions.
	}
	\label{fig:example}
\end{figure}

The magnon dispersion relation is identified to have a bandwidth of 20 meV and a 28 meV spin gap. We note that such a gapped magnon dispersion is consistent with the presence of antiferromagnetic all-in-all-out (AIAO) order. As discussed in Ref. \cite{Donnerer}, Donnerer et al. argued that the AIAO state is realized in Sm$_{2}$Ir$_{2}$O$_{7}$ based on the gapped spin wave dispersion and the polarization dependence of magnetic scattering. The magnon dispersion observed in Eu$_{2}$Ir$_{2}$O$_{7}$ is similar to that of Sm$_{2}$Ir$_{2}$O$_{7}$ with slightly larger gap size (28 meV for Eu$_{2}$Ir$_{2}$O$_{7}$ versus 25 meV in Sm$_{2}$Ir$_{2}$O$_{7}$), which strongly supports the AIAO ground state in Eu$_{2}$Ir$_{2}$O$_{7}$.

Next, we investigate how the magnetic order is destroyed at the phase transition, which provides an important clue as to the itinerancy of the system. On one extreme limit, in a completely itinerant picture, the magnetic order and metal-insulator transition occurs at the same time, and the magnetic fluctuation disappears above the transition temperature (similar to a Slater transition). On the other hand, in a Mott picture, the local magnetic moments will lose long range coherence, but they survive into the paramagnetic phase with their moments and interactions intact. In the latter case, the magnon at the zone boundary will be more or less unaffected by the magnetic transition, except for the Landau damping due to charge excitations.

\begin{figure*}
	\centerline{\includegraphics[width=6in,angle=0]{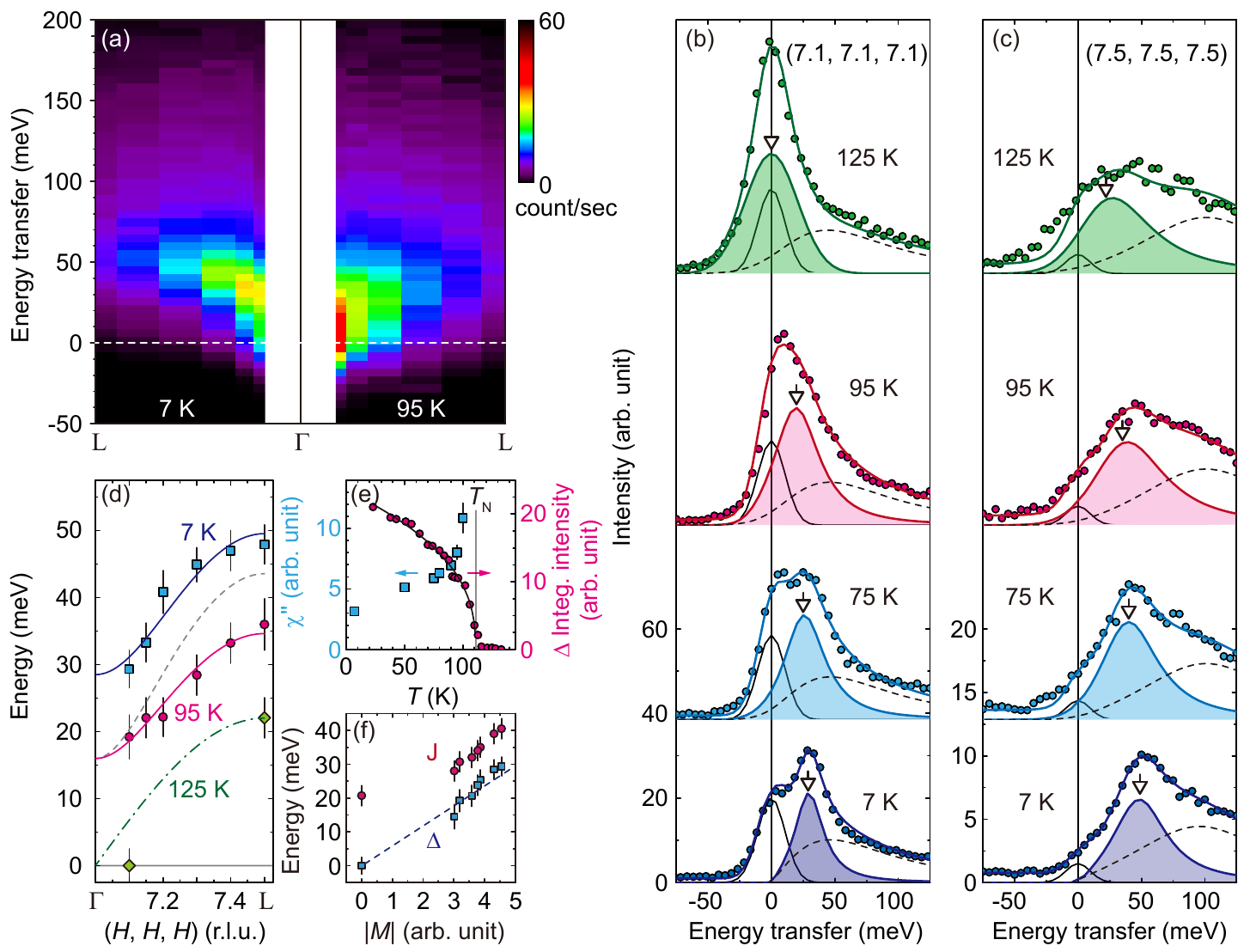}}
	\caption{		
		(a) Intensity maps of the RIXS spectra along the $\Gamma \rightarrow$ L line at \textit{T} = 7 K (left) and 95 K (right). Magnetic excitation monitored at \textit{\textbf{Q}} = (7.1, 7.1, 7.1) near the $\Gamma$ point (b) and at (7.5, 7.5, 7.5), the L point (c) for selected temperatures. The shaded peaks represent the damped harmonic oscillator model fitting of the magnetic excitations. The elastic peak and the temperature-independent (except for Bose factor) continuum background are denoted as the solid and dashed black lines, respectively. The $\omega_\textbf{\textit{q}}$ (open arrow) starts to be lower than the apparent position at high temperatures due to the Bose thermal factor. (d) Dispersive magnetic excitations along the $\Gamma \rightarrow$ L line at 7 K (square), 95 K (circle), and 125 K (rhombus). The solid lines are spin-wave fitting as described in the text. The grey and green dashed lines simulate magnon dispersion with ($\Delta$, J) = (17 meV, 40 meV) and (0 meV, 22meV), respectively. (e) Temperature dependence of the imaginary part of spin susceptibility $\chi''$ determined at (7.1, 7.1, 7.1) (square, left scale) and the integrated magnetic Bragg reflection intensity at (8, 6, 8) (circle, right scale). The solid line is a power law fit with $T_{N}$ = 111(7) K and $\beta$  = 0.20(1) deviating from 0.325 known for the 3D Heisenberg magnet \cite{Chen}. (f) Scaling of J (circle) and $\Delta$ (square) with the magnetic order parameter $|M|$. The dashed line is a guide to the eye.
	}\label{fig:fig3}
\end{figure*}

Figures 3(a,d) show the spectra along the $\Gamma$ $\rightarrow$ L line at \textit{T} = 7 K and 95 K, which is still below the metal-insulator transition temperature $T_{MI}$ = 111 K. Interestingly, we note that the magnon band in the Brillouin zone is rigidly shifted down to lower energy. The magnon energy decreases gradually as shown in the spectra at \textit{\textbf{Q}} = (7.1, 7.1, 7.1) and (7.5, 7.5, 7.5) for selected temperatures (see Figs. 3(b,c)). It is evident that the excitation near the zone center softens, and becomes gapless and highly damped above the transition temperature. The zone boundary excitation also softens, but still remains as a highly damped mode centered around $\sim$22 meV above the transition temperature (top spectrum at 125 K in Fig. 3(c)). This magnetic excitation dispersion can be modeled with a generic, antiferromagnetic spin wave expression, $\omega_q^{2} = \Delta^{2} + J^{2} $sin$^{2}$ $(\pi |\textit{\textbf{q}}|/\sqrt{3})$ \cite{Dyson} with substantially reduced ($\Delta$, J) as increasing temperature.

The collapse of the zone center gap energy $\Delta$ is expected on symmetry ground, since the system recovers rotational symmetry when the magnetic order vanishes. Figure 3(e) shows temperature dependence of the magnetic Bragg peak intensity at \textbf{\textit{Q}} = (8, 6, 8) that follows a power law, $I_{mag} \propto (T_{N}-T)^{2\beta}$ with $T_{N}$ = 111(7) K and $\beta$ = 0.20(1). We also observe that the imaginary part of magnetic susceptibility $\chi(\textit{\textbf{q}})''$ $\propto$ $A(\textit{\textbf{q}})$ of the spectrum near the zone center \textit{\textbf{q}} = (0.1, 0.1, 0.1) diverges as temperature approaches $T_{N}$. In addition, $\Delta$ shows a linear relation with the order parameter $|M|$ ($\propto \sqrt{I_{mag}}$), evidencing that the excitation gap scales with the order parameter (Fig. 3(f)). All these observations clearly demonstrate critical behavior expected for a 2nd order phase transition at $T_{N}$.

On the other hand, temperature dependence of J is surprising. If Eu$_{2}$Ir$_{2}$O$_{7}$ is a purely local moment system, the nearest neighbor exchange interactions are expected to change very little with temperature variation as mentioned above (the Ir-Ir bond length change by only $\sim$0.1 \% going from 7 K to 95 K). If we keep the same J, and use reduced $\Delta$ at 95 K, the spin wave dispersion predicts only about 10 \% softening of the zone boundary energy (the dashed line in Fig. 3(d)). In addition, the zone boundary magnon energy is renormalized due to magnon-magnon interaction. Physically, the excitation energy to flip a spin is lowered if the spin is already partially reversed due to thermal fluctuation. However, this renormalization effect is insignificant if the exchange energies are sufficiently larger than thermal energy scale. For example, La$_{2}$CuO$_{4}$, i.e. a parent compound of high temperature superconducting cuprates, shows only 4.5 \% decrease of the zone boundary energy at 295 K from the low temperature value \cite{Coldea}. This renormalization effect is even smaller for a three-dimensional material, and we estimate the zone boundary magnon renormalization in Eu$_{2}$Ir$_{2}$O$_{7}$ to be about 2.5 \% according to a Hartree-Fock calculation \cite{Davies}, which is clearly inadequate to account for almost 50 \% change observed in our study (at 125 K). Thus, our observation strongly suggests that Eu$_{2}$Ir$_{2}$O$_{7}$ cannot be described with a local moment model satisfactorily.

We note that Eu$_{2}$Ir$_{2}$O$_{7}$ undergoes a concomitant metal-insulator transition at $T_{N}$, and thus the role of charge fluctuation can be siginficant. As temperature rises, both spin and charge gaps soften, and both spin and charge fluctuations increase. In this system, the electrons close to the Fermi level will contribute to both magnetic moments and their exchange strengths. Therefore, the local moment picture is no longer applicable in such a situation, and the magnetism in Eu$_{2}$Ir$_{2}$O$_{7}$ requires an itinerant electron point of view. This description is consistent with the Pauli paramagnetic behavior deviating from the Curie-Weiss relation observed in the magnetic susceptibility above $T_{N}$ \cite{Ishikawa}. How charge fluctuations may affect the magnetism near $T_{N}$ requires quantitative calculation, which will be an intersting subject for future theoretical studies.

In summary, we have carried out Resonant Inelastic X-ray Scattering experiment to study the temperature dependence of the magnetic excitation spectrum in a pyrochlore iridate Eu$_{2}$Ir$_{2}$O$_{7}$. We observe a well-defined propagating magnon mode with with 20 meV bandwidth and a large gap of 28 meV at 7 K, which exhibits a drastic softening as temperature is raised toward $T_{N}$. The observed thermal renormalization behavior strongly indicates that the magnetism of Eu$_{2}$Ir$_{2}$O$_{7}$ is better described with itinerant electron formalism, suggesting electron correlation in this compound is not very strong.

Work at the University of Toronto was supported by the Natural Sciences and Engineering Research Council of Canada through Discovery and CREATE program, and Canada Foundation for Innovation. Use of the Advanced Photon Source at Argonne National Laboratory is supported by the US Department of Energy, Office of Science, under Contract No. DE-AC02-06CH11357. Use of the Canadian Light Source is supported by the Canada Foundation for Innovation, the Natural Sciences and Engineering Research Council of Canada, the University of Saskatchewan, the Government of Saskatchewan, Western Economic Diversification Canada, the National Research Council of Canada, and the Canadian Institutes of Health Research. Work at the University of Tokyo was supported by CREST, Japan Science and Technology Agency, Grants-in-Aid for Scientific Research (Grant Nos. 16H02209, 25707030), by Grants-in-Aid for Scientific Research on Innovative Areas ``J-Physics" (Grant Nos. 15H05882 and 15H05883) from the Japanese Society for the Promotion of Science.

\end{document}